\documentclass{article}

\usepackage{theorem}
\usepackage{latexsym}
\usepackage{amssymb}
\def\qed{\rule{1mm}{2.5mm}}

\theoremstyle{plain} \newtheorem{thm}{Theorem}[section]
\theoremstyle{plain} \newtheorem{prop}[thm]{Proposition}
\theoremstyle{plain} \newtheorem{lem}[thm]{Lemma}
\theoremstyle{plain} \newtheorem{rem}[thm]{Remark}
\theoremstyle{plain} 
\theoremstyle{plain} \newtheorem{df}[thm]{Definition}
\theoremstyle{plain} 

\makeatletter
  
  \@addtoreset{equation}{section}
\makeatother

\def\aa{{\alpha}}
\def\ep{{\varepsilon}}
\newcommand{\BZ}{{\mathbb Z}}

\begin{document}

\title{On the Rational Solutions of $q$-Painlev\'e V Equation}
\author{Tetsu Masuda \\
Department of Mathematics, Kobe University, \\
Rokko, Kobe, 657-8501, Japan \\
{\small masuda@math.kobe-u.ac.jp}}
\date{}

\maketitle


\begin{abstract}
We give an explicit determinant formula for a class of rational solutions of a $q$-analogue of the Painlev\'e V equation. 
The entries of the determinant are given by the continuous $q$-Laguerre polynomials. 
\end{abstract}

\section{Introduction\label{Intro}}
Since the introduction of the singularity confinement criterion as the discrete analogue of the Painlev\'e test~\cite{GRP}, a lot of ordinary difference equations have been proposed as discrete Painlev\'e equations~\cite{RGH,GR:physicaA}. 
It is known that the discrete Painlev\'e equations possess several properties analogous to the continuous ones such as the coalescence cascade, symmetry as the B\"acklund transformations and particular solutions. 

Recently, Kajiwara, Noumi and Yamada have proposed a $q$-analogue of Painlev\'e IV equation~\cite{q-P4}, and investigated the structure of symmetry and special solutions of the $q$-P$_{\rm IV}$. 
It has been shown that the $q$-P$_{\rm IV}$ admits two types of special solutions; one is the special function type solutions, which are expressed in terms of the continuous $q$-Hermite-Weber functions, and another is the rational solutions expressed as the ratio of a $q$-analogue of Okamoto polynomials~\cite{O3}. 

In this paper, we consider the symmetric form of $q$-P$_{\rm V}$
\begin{equation}
\begin{array}{c}
 \smallskip
 \displaystyle 
 \bar{a}_0=a_0, \quad \bar{a}_1=a_1, \quad \bar{a}_2=a_2, \quad \bar{a}_3=a_3, \\
 \smallskip
 \displaystyle 
 \bar{f}_0=a_0 a_1 f_1 
           \frac{1+a_2 f_2+a_2 a_3 f_2 f_3 +a_2 a_3 a_0 f_2 f_3 f_0}
                {1+a_0 f_0+a_0 a_1 f_0 f_1 +a_0 a_1 a_2 f_0 f_1 f_2}, \\ 
 \smallskip
 \displaystyle 
 \bar{f}_1=a_1 a_2 f_2 
           \frac{1+a_3 f_3+a_3 a_0 f_3 f_0 +a_3 a_0 a_1 f_3 f_0 f_1}
                {1+a_1 f_1+a_1 a_2 f_1 f_2 +a_1 a_2 a_3 f_1 f_2 f_3}, \\ 
 \smallskip
 \displaystyle 
 \bar{f}_2=a_2 a_3 f_3 
           \frac{1+a_0 f_0+a_0 a_1 f_0 f_1 +a_0 a_1 a_2 f_0 f_1 f_2}
                {1+a_2 f_2+a_2 a_3 f_2 f_3 +a_2 a_3 a_0 f_2 f_3 f_0}, \\ 
 \smallskip
 \displaystyle 
 \bar{f}_3=a_3 a_0 f_0 
           \frac{1+a_1 f_1+a_1 a_2 f_1 f_2 +a_1 a_2 a_3 f_1 f_2 f_3}
                {1+a_3 f_3+a_3 a_0 f_3 f_0 +a_3 a_0 a_1 f_3 f_0 f_1}, 
\end{array}   \label{q-P5}
\end{equation}
with 
\begin{equation}
a_0 a_1 a_2 a_3=q^{-1}, \label{norm:para}
\end{equation}
where $\bar{~}$ stands for the discrete time evolution. 
Introducing a variable $c$ by 
\begin{equation}
f_0 f_2=f_1 f_3=c^{-1}, \label{norm:f}
\end{equation}
we find that $c$ plays a role of the independent variable, 
\begin{equation}
\bar{c}=qc. \label{idv}
\end{equation}

Originally, the equation (\ref{q-P5}) is derived as a subsystem of the discrete dynamical systems associated with the extended affine Weyl group symmetry of type $A_{m-1}^{(1)}\times A_{n-1}^{(1)}$~\cite{KNY}. 
In the case of $(m,n)=(2,4)$, by regarding a translation of $\widetilde{W}(A_1^{(1)})$ as the discrete time evolution $\bar{~}$, we obtain the system (\ref{q-P5}). 
The variables $q$ and $c$ are invariant for the action of $\widetilde{W}(A_1^{(1)}\times A_3^{(1)})$ and $\widetilde{W}(A_3^{(1)})$, respectively. 
The inverse time evolution of (\ref{q-P5}) is given by 
\begin{equation}
\begin{array}{c}
 \displaystyle 
 \smallskip
  \underline{a_0}=a_0, \quad \underline{a_1}=a_1, \quad 
  \underline{a_2}=a_2, \quad \underline{a_3}=a_3, \\
 \smallskip
 \displaystyle 
  \underline{f_0}=\frac{f_3}{a_0 a_1}
                  \frac{a_2 a_1 a_0+a_1 a_0 f_2+a_0 f_2 f_1+f_2 f_1 f_0}
                       {a_0 a_3 a_2+a_3 a_2 f_0+a_2 f_0 f_3+f_0 f_3 f_2}, \\
 \smallskip
 \displaystyle 
  \underline{f_1}=\frac{f_0}{a_1 a_2}
                  \frac{a_3 a_2 a_1+a_2 a_1 f_3+a_1 f_3 f_2+f_3 f_2 f_1}
                       {a_1 a_0 a_3+a_0 a_3 f_1+a_3 f_1 f_0+f_1 f_0 f_3}, \\
 \smallskip
 \displaystyle 
  \underline{f_2}=\frac{f_1}{a_2 a_3}
                  \frac{a_0 a_3 a_2+a_3 a_2 f_0+a_2 f_0 f_3+f_0 f_3 f_2}
                       {a_2 a_1 a_0+a_1 a_0 f_2+a_0 f_2 f_1+f_2 f_1 f_0}, \\
 \smallskip
 \displaystyle 
  \underline{f_3}=\frac{f_0}{a_3 a_0}
                  \frac{a_1 a_0 a_3+a_0 a_3 f_1+a_3 f_1 f_0+f_1 f_0 f_3}
                       {a_3 a_2 a_1+a_2 a_1 f_3+a_1 f_3 f_2+f_3 f_2 f_1}. 
\end{array}
\end{equation}

The reason why we refer to the discrete system (\ref{q-P5}) as the symmetric form of $q$-P$_{\rm V}$ is as follows. 
By the original construction in \cite{KNY}, it is clear that this equation admits the affine Weyl group symmetry of type $A_3^{(1)}$ as the B\"acklund transformation group, which is stated in section \ref{sym:bi} precisely. 
Moreover, the system (\ref{q-P5}) reduces to the symmetric form of Painlev\'e V equation in the continuum limit. 
We set 
\begin{equation}
q=e^{\ep^2/2}, \quad 
a_i=e^{-\frac{\ep^2}{2}\aa_i}, \quad 
f_i=-e^{-\ep \varphi_i}, \quad 
c=e^{\ep \gamma}, 
\end{equation}
and define the derivation $\displaystyle \frac{d}{ds}$ by 
\begin{equation}
\frac{dz}{ds}=\lim_{\ep \to 0}\frac{\bar{z}-z}{\ep}, 
\end{equation}
for a function $z$ in $\aa_i$ and $\varphi_i$. 
Then, we get from (\ref{q-P5}) and (\ref{idv}) 
\begin{equation}
\frac{d\varphi_0}{ds}=\frac{1}{\gamma}
\left[\varphi_0 \varphi_2(\varphi_1-\varphi_3)
+\left(\frac{1}{2}-\aa_2\right)\varphi_0+\aa_0 \varphi_2 \right], 
\quad \frac{d\gamma}{ds}=\frac{1}{2}. 
\end{equation}
Introducing the variable $t$ and derivation $'$ as 
\begin{equation}
\gamma=\sqrt{t}, \quad '=t\frac{d}{dt},
\end{equation}
we have 
\begin{equation}
\begin{array}{c}
 \displaystyle 
  \aa_0'=0, \quad \aa_1'=0, \quad \aa_2'=0, \quad \aa_3'=0, \\
 \displaystyle 
  \varphi_0'=\varphi_0 \varphi_2(\varphi_1-\varphi_3)
   +\left(\frac{1}{2}-\aa_2\right)\varphi_0+\aa_0 \varphi_2, \\
 \displaystyle 
  \varphi_1'=\varphi_1 \varphi_3(\varphi_2-\varphi_0)
   +\left(\frac{1}{2}-\aa_3\right)\varphi_1+\aa_1 \varphi_3, \\
 \displaystyle 
  \varphi_2'=\varphi_2 \varphi_0(\varphi_3-\varphi_1)
   +\left(\frac{1}{2}-\aa_0\right)\varphi_2+\aa_2 \varphi_0, \\
 \displaystyle 
  \varphi_3'=\varphi_3 \varphi_1(\varphi_0-\varphi_2)
   +\left(\frac{1}{2}-\aa_1\right)\varphi_3+\aa_3 \varphi_1. 
\end{array}   \label{P5}
\end{equation}
The normalization conditions (\ref{norm:para}) and (\ref{norm:f}) reduce to 
\begin{equation}
\aa_0+\aa_1+\aa_2+\aa_3=1,
\end{equation}
and 
\begin{equation}
\varphi_0+\varphi_2=\varphi_1+\varphi_3=\sqrt{t}, \label{norm:phi}
\end{equation}
respectively. 
This differential system (\ref{P5})-(\ref{norm:phi}) is nothing but the symmetric form of P$_{\rm V}$~\cite{NY}. 

On the other hand, it has been revealed that a family of the rational solutions of P$_{\rm V}$, which exists on the barycenters of Weyl chambers, has a determinant formula whose entries are the Laguerre polynomials~\cite{p5:rat}. 
This determinant expression is regarded as a specialization of the universal characters~\cite{Koike}. 
The aim of this paper is to present an explicit determinant formula for a class of the rational solutions of $q$-P$_{\rm V}$. 

This paper is organized as follows.  
In Section \ref{result}, we give the main result of this paper. 
In Section \ref{sym:bi}, we describe the affine Weyl group symmetry and derive a set of bilinear equations for the $\tau$-functions of $q$-P$_{\rm V}$. 
In Section \ref{rat}, we construct the rational solutions of $q$-P$_{\rm V}$. 
Proof of our result is given in Section \ref{proof}. 
Section \ref{remarks} is devoted to some remarks.

\section{Main Result\label{result}}
\begin{df}
Let $p_k^{(b)}=p_k^{(b)}(y|q)$ and $q_k^{(b)}=q_k^{(b)}(y|q),~k \in \BZ$, be two sets of polynomials defined by 
\begin{equation}
 \begin{array}{l}
 \displaystyle 
  \sum_{k=0}^{\infty}p_k^{(b)}\lambda^k=
   \frac{(q^{\frac{1}{4}}b\lambda,q^{\frac{3}{4}}b\lambda;q)_{\infty}}
        {(-q^{\frac{1}{4}}x\lambda,-q^{\frac{3}{4}}x^{-1}\lambda;q)_{\infty}},
   \quad p_k^{(b)}=0\ \mbox{for}\ k<0, \\
 \displaystyle 
  \sum_{k=0}^{\infty}q_k^{(b)}\lambda^k=
   \frac{(-q^{\frac{1}{4}}x\lambda,-q^{\frac{3}{4}}x^{-1}\lambda;q)_{\infty}}
        {(q^{\frac{1}{4}}b^{-1}\lambda,q^{\frac{3}{4}}b^{-1}\lambda;q)_{\infty}},
   \quad q_k^{(b)}=0\ \mbox{for}\ k<0, 
 \end{array}\label{p-q:(+-)}
\end{equation}
with $\displaystyle y=-\frac{1}{2}\left(q^{-\frac{1}{4}}x+q^{\frac{1}{4}}x^{-1}\right)$. 
For $m,n \in \BZ_{\ge 0}$, we define a family of polynomials $R_{m,n}^{(b)}=R_{m,n}^{(b)}(y|q)$ by 
\begin{equation}
R_{m,n}^{(b)}=
 \left|
  \begin{array}{cccccccc}
   q_1^{(b)}        & q_0^{(b)}        & \cdots           & q_{-m+2}^{(b)}   &
   q_{-m+1}^{(b)}   & \cdots           & q_{-m-n+3}^{(b)} & q_{-m-n+2}^{(b)} \\
   q_3^{(b)}        & q_2^{(b)}        & \cdots           & q_{-m+4}^{(b)}   &
   q_{-m+3}^{(b)}   & \cdots           & q_{-m-n+5}^{(b)} & q_{-m-n+4}^{(b)} \\
   \vdots           & \vdots           & \ddots           & \vdots           &
   \vdots           & \ddots           & \vdots           & \vdots           \\
   q_{2m-1}^{(b)}   & q_{2m-2}^{(b)}   & \cdots           & q_m^{(b)}        &
   q_{m-1}^{(b)}    & \cdots           & q_{m-n+1}^{(b)}  & q_{m-n}^{(b)}    \\
   p_{n-m}^{(b)}    & p_{n-m+1}^{(b)}  & \cdots           & p_{n-1}^{(b)}    &
   p_n^{(b)}        & \cdots           & p_{2n-2}^{(b)}   & p_{2n-1}^{(b)}   \\
   \vdots           & \vdots           & \ddots           & \vdots           &
   \vdots           & \ddots           & \vdots           & \vdots           \\
   p_{-n-m+4}^{(b)} & p_{-n-m+5}^{(b)} & \cdots           & p_{-n+3}^{(b)}   &
   p_{-n+4}^{(b)}   & \cdots           & p_2^{(b)}        & p_3^{(b)}        \\
   p_{-n-m+2}^{(b)} & p_{-n-m+3}^{(b)} & \cdots           & p_{-n+1}^{(b)}   &
   p_{-n+2}^{(b)}   & \cdots           & p_0^{(b)}        & p_1^{(b)}
  \end{array}
 \right|. 
\end{equation}
For $m,n \in \BZ_{<0}$, we define $R_{m,n}^{(b)}$ through 
\begin{equation}
R_{m,n}^{(b)}=(-1)^{m(m+1)/2}R_{-m-1,n}^{(b)}, \quad 
R_{m,n}^{(b)}=(-1)^{n(n+1)/2}R_{m,-n-1}^{(b)}. \label{neg:R}
\end{equation}
\end{df}

\begin{rem}
The polynomials $p_k^{(b)}$ and $q_k^{(b)}$ are essentially the continuous $q$-Laguerre polynomials $P_k^{(\aa)}(y|q)$, which is defined by~\cite{KS} 
\begin{equation}
\sum_{k=0}^{\infty}P_k^{(\aa)}(y|q)\lambda^k=
 \frac{(q^{\aa+\frac{1}{2}}\lambda,q^{\aa+1}\lambda;q)_{\infty}}
      {(q^{\frac{1}{2}\aa+\frac{1}{4}}e^{ i \theta}\lambda,
        q^{\frac{1}{2}\aa+\frac{1}{4}}e^{-i \theta}\lambda;q)_{\infty}},
 \quad P_k^{(\aa)}(y|q)=0\ \mbox{for}\ k<0, 
 \quad y=\cos \theta. 
\end{equation}
In fact, denoting as 
\begin{equation}
L_k(y,b|q)=P_k^{(\aa)}(y|q), \quad b=q^{\frac{1}{2}\aa+\frac{1}{4}},
\end{equation}
we see that $p_k^{(b)}$ and $q_k^{(b)}$ are expressed as 
\begin{equation}
p_k^{(b)}(y|q)=(q^{\frac{3}{4}}b^{-1})^k L_k(y,q^{-\frac{1}{4}}b|q), \quad 
q_k^{(b)}(y|q)=(q^{-\frac{3}{4}}b)^k L_k(y,q^{\frac{1}{4}}b^{-1}|q^{-1}), 
\label{p-q:L}
\end{equation}
respectively. 
\end{rem}

Our main result is stated as follows. 

\begin{thm}\label{main}
We set 
\begin{equation}
S_{m,n}(x,a)=R_{m,n}^{(b)}(y), \quad 
y=-\frac{1}{2}\left(q^{-\frac{1}{4}}x+q^{\frac{1}{4}}x^{-1}\right), \quad 
b=q^{-\frac{1}{2}(m-n)}a. \label{yxba}
\end{equation}
Then, for the parameters 
\begin{equation}
(a_0,a_1,a_2,a_3)=
\left(q^{n-\frac{1}{2}}a,a^{-1},q^{-m-\frac{1}{2}}a,q^{m-n}a^{-1}\right), 
\end{equation}
we have the following rational solutions of $q$-P$_{\rm V}$, 
\begin{equation}
\begin{array}{l}
 \smallskip
 \displaystyle 
  1+q^{\frac{1}{2}(2n-1)}a f_0(x,a)=
   q^{\frac{1}{2}m}(1+q^{-\frac{1}{2}(m-n+1)}ax^{-1})
   \frac{S_{m,n}(x,a)S_{m-1,n-1}(q^{\frac{1}{2}}x,q^{-1}a)}
        {S_{m,n-1}(q^{\frac{1}{2}}x,a)S_{m-1,n}(x,q^{-1}a)}, \\
 \smallskip
 \displaystyle 
  1+q^{-\frac{1}{2}(2n-1)}a^{-1}f_0(x,a)=
   q^{-\frac{1}{2}n}(1+q^{\frac{1}{2}(m-n+1)}a^{-1}x^{-1})
   \frac{S_{m,n}(q^{\frac{1}{2}}x,a)S_{m-1,n-1}(x,q^{-1}a)}
        {S_{m,n-1}(q^{\frac{1}{2}}x,a)S_{m-1,n}(x,q^{-1}a)}, \\
 \smallskip
 \displaystyle 
  1+a^{-1}f_1(x,a)=
   q^{\frac{1}{2}n}(1+q^{\frac{1}{2}(m-n)}a^{-1}x^{-1})
   \frac{S_{m-1,n}(x,q^{-1}a)S_{m,n-1}(q^{\frac{1}{2}}x,qa)}
        {S_{m,n}(q^{\frac{1}{2}}x,a)S_{m-1,n-1}(x,a)}, \\
 \smallskip
 \displaystyle 
  1+a f_1(x,a)=
   q^{\frac{1}{2}m}(1+q^{-\frac{1}{2}(m-n)}ax^{-1})
   \frac{S_{m-1,n}(q^{\frac{1}{2}}x,q^{-1}a)S_{m,n-1}(x,qa)}
        {S_{m,n}(q^{\frac{1}{2}}x,a)S_{m-1,n-1}(x,a)}, \\
 \smallskip
 \displaystyle 
  1+q^{-\frac{1}{2}(2m+1)}a f_2(x,a)=
   q^{-\frac{1}{2}m}(1+q^{-\frac{1}{2}(m-n+1)}ax^{-1})
   \frac{S_{m-1,n-1}(x,a)S_{m,n}(q^{\frac{1}{2}}x,q^{-1}a)}
        {S_{m-1,n}(q^{\frac{1}{2}}x,q^{-1}a)S_{m,n-1}(x,a)}, \\
 \smallskip
 \displaystyle 
  1+q^{\frac{1}{2}(2m+1)}a^{-1}f_2(x,a)=
   q^{\frac{1}{2}n}(1+q^{\frac{1}{2}(m-n+1)}a^{-1}x^{-1})
   \frac{S_{m-1,n-1}(q^{\frac{1}{2}}x,a)S_{m,n}(x,q^{-1}a)}
        {S_{m-1,n}(q^{\frac{1}{2}}x,q^{-1}a)S_{m,n-1}(x,a)}, \\
 \smallskip
 \displaystyle 
  1+q^{m-n}a^{-1}f_3(x,a)=
   q^{-\frac{1}{2}n}(1+q^{\frac{1}{2}(m-n)}a^{-1}x^{-1})
   \frac{S_{m,n-1}(x,a)S_{m-1,n}(q^{\frac{1}{2}}x,a)}
        {S_{m-1,n-1}(q^{\frac{1}{2}}x,a)S_{m,n}(x,a)}, \\
 \smallskip
 \displaystyle 
  1+q^{-m+n}a f_3(x,a)=
   q^{-\frac{1}{2}m}(1+q^{-\frac{1}{2}(m-n)}ax^{-1})
   \frac{S_{m,n-1}(q^{\frac{1}{2}}x,a)S_{m-1,n}(x,a)}
        {S_{m-1,n-1}(q^{\frac{1}{2}}x,a)S_{m,n}(x,a)},
\end{array}   \label{f-S}
\end{equation}
with $x^2=c$. 
\end{thm}

\section{Weyl Group Symmetry and Bilinear Relations \label{sym:bi}}
As we mentioned in Section \ref{Intro}, the symmetric form of $q$-P$_{\rm V}$ (\ref{q-P5}) admits the symmetry of the extended affine Weyl group $\widetilde{W}=<s_0,s_1,s_2,s_3,\pi>$ of type $A_3^{(1)}$ as a group of B\"acklund transformations. 
The action of $s_i$ and $\pi$ on the variables $a_i$ and $f_i$ is given by 
\begin{equation}
s_i(a_j)=a_j a_i^{-a_{ij}}, \quad \pi(a_j)=a_{j+1},
\end{equation}
\begin{equation}
s_i(f_j)=f_j \left(\frac{a_i+f_i}{1+a_i f_i}\right)^{u_{ij}}, \quad 
\pi(f_j)=f_{j+1}, \label{BT:f}
\end{equation}
where $A=(a_{ij})_{i,j=0}^3$ is the generalized Cartan matrix of type $A_3^{(1)}$ and $U=(u_{ij})_{i,j=0}^3$ is an orientation matrix of the corresponding Dynkin diagram 
\begin{equation}
A=
 \left(
  \begin{array}{cccc}
    2 & -1 &  0 & -1 \\
   -1 &  2 & -1 &  0 \\
    0 & -1 &  2 & -1 \\
   -1 &  0 & -1 &  2
  \end{array}
 \right),    \quad 
U=
 \left(
  \begin{array}{cccc}
    0 &  1 &  0 & -1 \\
   -1 &  0 &  1 &  0 \\
    0 & -1 &  0 &  1 \\
    1 &  0 & -1 &  0
  \end{array}
 \right),    \label{AU}
\end{equation}
and indecies are understood as elements of $\BZ/4\BZ$. 
These transformations commute with the time evolution and satisfy the fundamental relations 
\begin{equation}
 \left.
  \begin{array}{c}
  \displaystyle 
   s_i^2=1, \quad s_i s_j =s_j s_i~(j \ne i,i \pm 1), \quad 
   s_i s_j s_i=s_j s_i s_j~(j=i \pm 1), \\
  \displaystyle 
   \pi^4=1, \quad \pi s_j=s_{j+1}\pi. 
 \end{array}
\right.     \label{fun.rel}
\end{equation}

Let us introduce $\tau$-functions $\tau_i$ as solutions of the following equations~\cite{private}, 
\begin{equation}
\bar{\bar{\tau_i}}=g_i\frac{\bar{\tau}_i \bar{\tau}_{i+1}}{\tau_{i+1}}, 
\end{equation}
where $g_i$ is given by 
\begin{equation}
g_i=1+a_{i+1}f_{i+1}+a_{i+1}a_{i+2}f_{i+1}f_{i+2}
     +a_{i+1}a_{i+2}a_{i+3}f_{i+1}f_{i+2}f_{i+3}. 
\end{equation}
The inverse transformations are given as, 
\begin{equation}
\underline{\tau_i}=h_i\frac{\tau_{i-1}\tau_i}{\bar{\tau}_{i-1}}, 
\end{equation}
with 
\begin{equation}
h_i=1+\frac{f_{i-1}}{a_{i-1}}+\frac{f_{i-1}f_{i-2}}{a_{i-1}a_{i-2}}
     +\frac{f_{i-1}f_{i-2}f_{i-3}}{a_{i-1}a_{i-2}a_{i-3}}. 
\end{equation}
The B\"acklund transformations can be lifted on the $\tau$-functions as follows:
\begin{equation}
 \begin{array}{c}
  \displaystyle 
  s_i(\tau_i)=\left(1+\frac{f_i}{a_i}\right)
   \frac{\bar{\tau}_{i-1}\tau_{i+1}}{\bar{\tau}_i}, \quad 
  s_i(\bar{\tau}_i)=(1+a_i f_i)
   \frac{\bar{\tau}_{i-1}\tau_{i+1}}{\tau_i}, \\
  s_i(\tau_j)=\tau_j, \quad s_i(\bar{\tau}_j)=\bar{\tau}_j, \quad (i\ne j), \\
  \pi(\tau_j)=\tau_{j+1}, \quad \pi(\bar{\tau}_j)=\bar{\tau}_{j+1}. 
 \end{array} \label{BT:tau}
\end{equation}
The fundamental relations (\ref{fun.rel}) are preserved in this lifting.
Note that we have the multiplicative formulas 
\begin{equation}
1+\frac{f_i}{a_i}=\frac{\bar{\tau}_i s_i(\tau_i)}{\bar{\tau}_{i-1}\tau_{i+1}}, \quad
1+a_i f_i=\frac{\tau_i s_i(\bar{\tau}_i)}{\bar{\tau}_{i-1}\tau_{i+1}}, 
\label{multi:form}
\end{equation}
for the independent variables $f_i$ in terms of $\tau$-functions. 

\begin{prop}
We have the following bilinear equations: 
\begin{equation}
 \begin{array}{l}
 \smallskip
  \tau_0 s_0s_1(\bar{\tau}_1)=
   a_0^2 s_0(\tau_0)s_1(\bar{\tau}_1)+(1-a_0^2)\tau_2 \bar{\tau}_3, \\
 \smallskip
  \bar{\tau}_1 s_1s_0(\tau_0)=
   a_1^{-2}s_0(\tau_0)s_1(\bar{\tau}_1)+(1-a_1^{-2})\tau_2 \bar{\tau}_3, \\
 \smallskip
  \tau_1 s_1s_2(\bar{\tau}_2)=
   a_1^2 s_1(\tau_1)s_2(\bar{\tau}_2)+(1-a_1^2)\tau_3 \bar{\tau}_0, \\
 \smallskip
  \bar{\tau}_2 s_2s_1(\tau_1)=
   a_2^{-2}s_1(\tau_1)s_2(\bar{\tau}_2)+(1-a_2^{-2})\tau_3 \bar{\tau}_0, \\
 \smallskip
  \tau_2 s_2s_3(\bar{\tau}_3)=
   a_2^2 s_2(\tau_2)s_3(\bar{\tau}_3)+(1-a_2^2)\tau_3 \bar{\tau}_1, \\
 \smallskip
  \bar{\tau}_3 s_3s_2(\tau_2)=
   a_3^{-2}s_2(\tau_2)s_3(\bar{\tau}_3)+(1-a_3^{-2})\tau_0 \bar{\tau}_1, \\
 \smallskip
  \tau_3 s_3s_0(\bar{\tau}_0)=
   a_3^2 s_3(\tau_3)s_0(\bar{\tau}_0)+(1-a_3^2)\tau_1 \bar{\tau}_2, \\
 \smallskip
  \bar{\tau}_0 s_0s_3(\tau_3)=
   a_0^{-2}s_3(\tau_3)s_0(\bar{\tau}_0)+(1-a_0^{-2})\tau_1 \bar{\tau}_2. 
 \end{array} \label{bi:bt}
\end{equation}
\end{prop}

\noindent
{\it Proof.}~Eliminating $f_0$ from (\ref{BT:tau}) with $i=0$, we obtain 
\begin{equation}
1-a_0^2\frac{\bar{\tau}_0 s_0(\tau_0)}{\tau_0 s_0(\bar{\tau}_0)}=
(1-a_0^2)\frac{\bar{\tau}_3 \tau_1}{\tau_0 s_0(\bar{\tau}_0)}.  \label{aux}
\end{equation}
From (\ref{BT:f}) and (\ref{BT:tau}), we get 
\begin{equation}
\frac{\tau_1s_0s_1(\bar{\tau}_1)}{s_0(\bar{\tau}_0)\tau_2}=
1+a_0^2\left(\frac{\tau_1 s_1(\bar{\tau}_1)}{\bar{\tau}_0\tau_2}-1\right)
       \frac{\bar{\tau}_0 s_0(\tau_0)}{\tau_0 s_0(\bar{\tau}_0)}, 
\end{equation}
which leads to the first equation of (\ref{bi:bt}) by using (\ref{aux}). 
The other equations are derived by the similar way. 
\hfill \qed

\medskip

Let us define the translation operators $T_i~(i=0,1,2,3)$ by 
\begin{equation}
T_1=\pi s_3s_2s_1, \quad T_2=s_1\pi s_3s_2, \quad 
T_3=s_2s_1\pi s_3, \quad T_0=s_3s_2s_1\pi,       \label{def:T}
\end{equation}
which commute with each other and satisfy $T_1T_2T_3T_0=1$. 
These operators act on parameters $a_i$ as 
\begin{equation}
T_i(a_{i-1})=q^{-1}a_{i-1}, \quad T_i(a_i)=q a_i, \quad 
T_i(a_j)=a_j~(j \ne i-1,i). 
\end{equation}
In terms of $T_i$, $\tau$-functions in (\ref{bi:bt}) are expressed as 
\begin{equation}
  \begin{array}{lll}
\smallskip
   {\displaystyle \tau_1=T_1(\tau_0),}&{\displaystyle \tau_2=T_1T_2(\tau_0)}, &
{\displaystyle \tau_3=T_0^{-1}(\tau_0),} \\
\smallskip
   {\displaystyle s_0(\tau_0)=T_0^{-1}T_1(\tau_0),}&{\displaystyle s_1(\tau_1)=T_2(\tau_0),}&
  {\displaystyle s_2(\tau_2)=T_1T_3(\tau_0),}\\
\smallskip
{\displaystyle s_3(\tau_3)=T_3^{-1}(\tau_0), } &{\displaystyle s_0s_1(\tau_1)=T_1T_2T_0^{-1}(\tau_0),}
&{\displaystyle s_1s_0(\tau_0)=T_2T_0^{-1}(\tau_0),}\\
\smallskip
{\displaystyle s_1s_2(\tau_2)=T_2T_3(\tau_0),}&{\displaystyle s_2s_1(\tau_1)=T_3(\tau_0), }
&{\displaystyle s_2s_3(\tau_3)=T_2^{-1}(\tau_0),}\\
\smallskip
{\displaystyle s_3s_2(\tau_2)=T_1T_0(\tau_0), } & {\displaystyle s_3s_0(\tau_0)=T_1T_3^{-1}(\tau_0),}
&{\displaystyle s_0s_3(\tau_3)=T_1T_3^{-1}T_0^{-1}(\tau_0)}.
 \end{array}
\end{equation}
Introducing a notation, 
\begin{equation}
\tau_{l,m,n}=T_2^lT_3^mT_0^n(\tau_0), \quad 
\bar{\tau}_{l,m,n}=T_2^lT_3^mT_0^n(\bar{\tau}_0), \quad l,m,n \in \BZ, 
\end{equation}
we can express the bilinear relations (\ref{bi:bt}) as 
\begin{equation}
 \begin{array}{l}
 \smallskip
  \displaystyle 
   \tau_{l,m,n}\bar{\tau}_{l,m-1,n-2}=
   a_0^2 q^{2n}\tau_{l-1,m-1,n-2}\bar{\tau}_{l+1,m,n}
    +(1-a_0^2 q^{2n})\tau_{l,m-1,n-1}\bar{\tau}_{l,m,n-1}, \\
 \smallskip
  \displaystyle 
   \bar{\tau}_{l-1,m-1,n-1}\tau_{l+1,m,n-1}=
   a_1^{-2}q^{2l}\tau_{l-1,m-1,n-2}\bar{\tau}_{l+1,m,n}
    +(1-a_1^{-2}q^{2l})\tau_{l,m-1,n-1}\bar{\tau}_{l,m,n-1}, \\
 \smallskip
  \displaystyle 
   \tau_{l-1,m-1,n-1}\bar{\tau}_{l+1,m+1,n}=
   a_1^2 q^{-2l}\tau_{l+1,m,n}\bar{\tau}_{l-1,m,n-1}
    +(1-a_1^2 q^{-2l})\tau_{l,m,n-1}\bar{\tau}_{l,m,n}, \\
 \smallskip
  \displaystyle 
   \bar{\tau}_{l,m-1,n-1}\tau_{l,m+1,n}=
   a_2^{-2}q^{2(-l+m)}\tau_{l+1,m,n}\bar{\tau}_{l-1,m,n-1}
    +(1-a_2^{-2}q^{2(-l+m)})\tau_{l,m,n-1}\bar{\tau}_{l,m,n}, \\
 \smallskip
  \displaystyle 
   \tau_{l,m-1,n-1}\bar{\tau}_{l-1,m,n}=
   a_2^2 q^{2(l-m)}\tau_{l-1,m,n-1}\bar{\tau}_{l,m-1,n}
    +(1-a_2^2 q^{2(l-m)})\tau_{l,m,n}\bar{\tau}_{l-1,m-1,n-1}, \\
 \smallskip
  \displaystyle 
   \bar{\tau}_{l,m,n-1}\tau_{l-1,m-1,n}=
   a_3^{-2}q^{2(-m+n)}\tau_{l-1,m,n-1}\bar{\tau}_{l,m-1,n}
    +(1-a_3^{-2}q^{2(-m+n)})\tau_{l,m,n}\bar{\tau}_{l-1,m-1,n-1}, \\
 \smallskip
  \displaystyle 
   \tau_{l,m,n-1}\bar{\tau}_{l-1,m-2,n-1}=
   a_3^2 q^{2(m-n)}\tau_{l,m-1,n}\bar{\tau}_{l-1,m-1,n-2}
    +(1-a_3^2 q^{2(m-n)})\tau_{l-1,m-1,n-1}\bar{\tau}_{l,m-1,n-1}, \\
 \smallskip
  \displaystyle 
   \bar{\tau}_{l,m,n}\tau_{l-1,m-2,n-2}=
   a_0^{-2}q^{-2n}\tau_{l,m-1,n}\bar{\tau}_{l-1,m-1,n-2}
    +(1-a_0^{-2}q^{-2n})\tau_{l-1,m-1,n-1}\bar{\tau}_{l,m-1,n-1}. 
 \end{array}  \label{bi:bt:tau}
\end{equation}

\section{Construction of Rational Solutions \label{rat}}
In this section, we construct a family of rational solutions of $q$-P$_{\rm V}$. 
Similarly to the continuous case, we consider the fixed points of Dynkin diagram automorphism $\pi^2$ to get the seed solution. 
It is clear that the symmetric form of $q$-P$_{\rm V}$ (\ref{q-P5}) has a particular solution, 
\begin{equation}
(a_0,a_1,a_2,a_3)=
\left(q^{-\frac{1}{2}}a,a^{-1},q^{-\frac{1}{2}}a,a^{-1}\right), 
\quad (f_0,f_1,f_2,f_3)=(x^{-1},x^{-1},x^{-1},x^{-1}), \quad x^2=c, 
\label{seed}
\end{equation}
Applying B\"acklund transformations to the seed solution (\ref{seed}), we obtain a family of rational solutions of $q$-P$_{\rm V}$.

Calculating $\tau_{l,m,n}$ and $\bar{\tau}_{l,m,n}$ from (\ref{BT:f}),(\ref{BT:tau}) and (\ref{def:T}), and putting as (\ref{seed}) and 
\begin{equation}
\tau_i=\bar{\tau}_i=1, 
\end{equation}
we obtain the $\tau$-functions for the rational solutions of $q$-P$_{\rm V}$. 
For small $l,m,n$, we observe that $\tau_{l,m,n}$ and $\bar{\tau}_{l,m,n}$ are factorized as the form of 
\begin{equation}
\tau_{l,m,n}=c_k U_{l,m,n}, \quad 
\bar{\tau}_{l,m,n}=\tilde{c}_k \bar{U}_{l,m,n}, \quad k=m-n-l.  \label{tau-U}
\end{equation}
It is possible to guess that $U_{l,m,n}=U_{l,m,n}(x,a)$ are some polynomials in $x^{-1},a^{\pm 1}$ and $q^{\pm \frac{1}{2}}$, and that the factors $c_k$ and $\tilde{c}_k$ are determined by the recurrence relations 
\begin{equation}
c_{k+1}\tilde{c}_{k-1}=
\left(1+q^{\frac{k}{2}}a^{-1}x^{-1}\right)c_k \tilde{c}_k,  \quad 
\tilde{c}_{k+1}c_{k-1}=
\left(1+q^{-\frac{k}{2}}ax^{-1}\right)c_k \tilde{c}_k, 
\end{equation}
with the initial conditions 
\begin{equation}
c_0=c_1=1, \quad \tilde{c}_0=\tilde{c}_1=1. 
\end{equation}
Some examples of them are listed in Appendix \ref{example}. 

Notice that we have 
\begin{equation}
T_2^l T_0^l (a_0,a_1,a_2,a_3)=
\left(q^{-\frac{1}{2}}\tilde{a},\tilde{a}^{-1},q^{-\frac{1}{2}}\tilde{a},\tilde{a}^{-1}\right),  \quad \tilde{a}=q^l a, \quad l \in \BZ, \label{seed'}
\end{equation}
under the specialization of (\ref{seed}). 
Comparing (\ref{seed'}) with (\ref{seed}), we see that the effect of $T_2$ is absorbed by that of $T_0^{-1}$ and rescaling of the parameter $a$. 
Then, we do not need to consider the translation $T_2$ for constructing the family of rational solutions of $q$-P$_{\rm V}$, and it is possible to put  
\begin{equation}
U_{l,m,n}(x,a)=U_{0,m,n-l}(x,q^la). \label{sym:U}
\end{equation}

Now, by (\ref{seed}),(\ref{tau-U}) and (\ref{sym:U}), we can rewrite the bilinear relations (\ref{bi:bt:tau}) in terms of $U_{m,n}=U_{0,m,n}$. 
We have 
\begin{equation}
 \begin{array}{l}
 \smallskip
  \displaystyle 
   U_{m,n}\bar{U}_{m-1,n-2}=
   a^2 q^{2n-1}\mu_{m-n} U_{m-1,n-1}^{--}\bar{U}_{m,n-1}^{++}
    +(1-a^2 q^{2n-1})U_{m-1,n-1}\bar{U}_{m,n-1}, \\
 \smallskip
  \displaystyle 
   \bar{U}_{m-1,n}^{--}U_{m,n-2}^{++}=
   a^2\mu_{m-n} U_{m-1,n-1}^{--}\bar{U}_{m,n-1}^{++}
    +(1-a^2)U_{m-1,n-1}\bar{U}_{m,n-1}, \\
 \smallskip
  \displaystyle 
   U_{m-1,n}^{--}\bar{U}_{m+1,n-1}^{++}=
   a^{-2}\nu_{m-n} U_{m,n-1}^{++}\bar{U}_{m,n}^{--}
    +(1-a^{-2})U_{m,n-1}\bar{U}_{m,n}, \\
 \smallskip
  \displaystyle 
   \bar{U}_{m-1,n-1}U_{m+1,n}=
   a^{-2}q^{2m+1}\nu_{m-n} U_{m,n-1}^{++}\bar{U}_{m,n}^{--}
    +(1-a^{-2}q^{2m+1})U_{m,n-1}\bar{U}_{m,n}, \\
 \smallskip
  \displaystyle 
   U_{m-1,n-1}\bar{U}_{m,n+1}^{--}=
   a^2 q^{-2m-1}\mu_{m-n} U_{m,n}^{--}\bar{U}_{m-1,n}
    +(1-a^2 q^{-2m-1})U_{m,n}\bar{U}_{m-1,n}^{--}, \\
 \smallskip
  \displaystyle 
   \bar{U}_{m,n-1}U_{m-1,n+1}^{--}=
   a^2 q^{-2(m-n)}\mu_{m-n} U_{m,n}^{--}\bar{U}_{m-1,n}
    +(1-a^2 q^{-2(m-n)})U_{m,n}\bar{U}_{m-1,n}^{--}, \\
 \smallskip
  \displaystyle 
   U_{m,n-1}\bar{U}_{m-2,n}^{--}=
   a^{-2}q^{2(m-n)}\nu_{m-n} U_{m-1,n}\bar{U}_{m-1,n-1}^{--}
    +(1-a^{-2}q^{2(m-n)})U_{m-1,n}^{--}\bar{U}_{m-1,n-1}, \\
 \smallskip
  \displaystyle 
   \bar{U}_{m,n}U_{m-2,n-1}^{--}=
   a^{-2}q^{-2n+1}\nu_{m-n} U_{m-1,n}\bar{U}_{m-1,n-1}^{--}
    +(1-a^{-2}q^{-2n+1})U_{m-1,n}^{--}\bar{U}_{m-1,n-1},
 \end{array}  \label{bi:U}
\end{equation}
with the initial conditions
\begin{equation}
U_{-1,-1}=U_{-1,0}=U_{0,-1}=U_{0,0}=1, \label{init}
\end{equation}
where we denote as 
\begin{equation}
\mu_k=(1+q^{\frac{1}{2}(k+1)}a^{-1}x^{-1})(1+q^{\frac{k}{2}}a^{-1}x^{-1}),\quad 
\nu_k=(1+q^{-\frac{1}{2}(k-1)}ax^{-1})(1+q^{-\frac{k}{2}}ax^{-1}), 
\end{equation}
and 
\begin{equation}
U_{m,n}^{\pm \pm}=U_{m,n}(x,q^{\pm 1}a). 
\end{equation}

Conversely, by solving the bilinear relations (\ref{bi:U}) with (\ref{init}), we can construct the family of rational solutions of $q$-P$_{\rm V}$. 
Applying $T_3^mT_0^n$ to (\ref{multi:form}) and denoting as $T_3^mT_0^n(f_i)=f_i(x,a)$, we have the following Proposition. 

\begin{prop}\label{f:U}
Let $U_{m,n}=U_{m,n}(x,a)~(m,n \in \BZ)$ be polynomials in $x^{-1},a^{\pm 1}$ and $q^{\pm \frac{1}{2}}$ which satisfy the bilinear equations (\ref{bi:U}) with the initial conditions (\ref{init}). 
Then, $f_i(x,a)$ given by 
\begin{equation}
\begin{array}{l}
 \smallskip
 \displaystyle 
  1+q^{\frac{1}{2}(2n-1)}a f_0(x,a)=
   (1+q^{-\frac{1}{2}(m-n+1)}ax^{-1})
   \frac{U_{m,n}(x,a)U_{m-1,n-1}(q^{\frac{1}{2}}x,q^{-1}a)}
        {U_{m,n-1}(q^{\frac{1}{2}}x,a)U_{m-1,n}(x,q^{-1}a)}, \\
 \smallskip
 \displaystyle 
  1+q^{-\frac{1}{2}(2n-1)}a^{-1}f_0(x,a)=
   (1+q^{\frac{1}{2}(m-n+1)}a^{-1}x^{-1})
   \frac{U_{m,n}(q^{\frac{1}{2}}x,a)U_{m-1,n-1}(x,q^{-1}a)}
        {U_{m,n-1}(q^{\frac{1}{2}}x,a)U_{m-1,n}(x,q^{-1}a)}, \\
 \smallskip
 \displaystyle 
  1+a^{-1}f_1(x,a)=
   (1+q^{\frac{1}{2}(m-n)}a^{-1}x^{-1})
   \frac{U_{m-1,n}(x,q^{-1}a)U_{m,n-1}(q^{\frac{1}{2}}x,qa)}
        {U_{m,n}(q^{\frac{1}{2}}x,a)U_{m-1,n-1}(x,a)}, \\
 \smallskip
 \displaystyle 
  1+a f_1(x,a)=
   (1+q^{-\frac{1}{2}(m-n)}ax^{-1})
   \frac{U_{m-1,n}(q^{\frac{1}{2}}x,q^{-1}a)U_{m,n-1}(x,qa)}
        {U_{m,n}(q^{\frac{1}{2}}x,a)U_{m-1,n-1}(x,a)}, \\
 \smallskip
 \displaystyle 
  1+q^{-\frac{1}{2}(2m+1)}a f_2(x,a)=
   (1+q^{-\frac{1}{2}(m-n+1)}ax^{-1})
   \frac{U_{m-1,n-1}(x,a)U_{m,n}(q^{\frac{1}{2}}x,q^{-1}a)}
        {U_{m-1,n}(q^{\frac{1}{2}}x,q^{-1}a)U_{m,n-1}(x,a)}, \\
 \smallskip
 \displaystyle 
  1+q^{\frac{1}{2}(2m+1)}a^{-1}f_2(x,a)=
   (1+q^{\frac{1}{2}(m-n+1)}a^{-1}x^{-1})
   \frac{U_{m-1,n-1}(q^{\frac{1}{2}}x,a)U_{m,n}(x,q^{-1}a)}
        {U_{m-1,n}(q^{\frac{1}{2}}x,q^{-1}a)U_{m,n-1}(x,a)}, \\
 \smallskip
 \displaystyle 
  1+q^{m-n}a^{-1}f_3(x,a)=
   (1+q^{\frac{1}{2}(m-n)}a^{-1}x^{-1})
   \frac{U_{m,n-1}(x,a)U_{m-1,n}(q^{\frac{1}{2}}x,a)}
        {U_{m-1,n-1}(q^{\frac{1}{2}}x,a)U_{m,n}(x,a)}, \\
 \smallskip
 \displaystyle 
  1+q^{-m+n}a f_3(x,a)=
   (1+q^{-\frac{1}{2}(m-n)}ax^{-1})
   \frac{U_{m,n-1}(q^{\frac{1}{2}}x,a)U_{m-1,n}(x,a)}
        {U_{m-1,n-1}(q^{\frac{1}{2}}x,a)U_{m,n}(x,a)},
\end{array}   \label{f-U}
\end{equation}
solve the $q$-P$_{\rm V}$ (\ref{q-P5}) for the parameters 
\begin{equation}
(a_0,a_1,a_2,a_3)=
\left(q^{n-\frac{1}{2}}a,a^{-1},q^{-m-\frac{1}{2}}a,q^{m-n}a^{-1}\right). 
\end{equation}
\end{prop}

\section{Proof of Theorem \ref{main} \label{proof}}
In this section, we give the proof for Theorem \ref{main}.

\begin{prop}\label{U:S}
We have 
\begin{equation}
U_{m,n}=(-1)^{\left({{n+1}\atop{2}}\right)}\kappa_m \kappa_n S_{m,n},  \label{U-S}
\end{equation}
where $S_{m,n}=S_{m,n}(x,a)$ is defined in Theorem \ref{main} and $\kappa_n$ is the factor determined by 
\begin{equation}
\kappa_{n+1}\bar{\kappa}_{n-1}=q^{-\frac{1}{4}(2n+1)}x^{-1}(1-q^{2n+1})\kappa_n \bar{\kappa}_n, 
\quad \kappa_0=\kappa_{-1}=1. 
\end{equation}
\end{prop}
We notice that $\kappa_n$ for $n \ge 0$ is expressed as 
\begin{equation}
\kappa_n=
q^{-\left({{n+1}\atop{3}}\right)-\frac{1}{4}\left({{n+1}\atop{2}}\right)}
x^{-\left({{n+1}\atop{2}}\right)}\prod_{k=1}^n \prod_{j=1}^k (1-q^{2j-1}). 
\end{equation}

By substituting (\ref{U-S}) into (\ref{f-U}), we find that Theorem \ref{main} is a direct consequence of Proposition \ref{f:U} and \ref{U:S}. 
Taking (\ref{U-S}) and (\ref{yxba}) into account, we obtain the bilinear relations for $R_{m,n}^{(b)}$. 

\begin{prop}\label{bi:rel}
The following bilinear relations hold: 
\begin{equation}
 \begin{array}{l}
 \smallskip
  \displaystyle 
  -q^{\frac{1}{4}(2m-2n-1)}x(1-q^{2n+1})R_{m,n+1}^{+}\bar{R}_{m-1,n-1} \\
 \smallskip
  \displaystyle 
   \hskip50pt =b^2 q^{m+n+1}\mu^{+}R_{m-1,n}^{-}\bar{R}_{m,n}^{++}
    +x^2(1-b^2 q^{m+n+1})R_{m-1,n}^{+}\bar{R}_{m,n}, \\
 \smallskip
  \displaystyle 
  -q^{\frac{1}{4}(2m-6n-3)}x(1-q^{2n+1})\bar{R}_{m-1,n+1}R_{m,n-1}^{+} \\
 \smallskip
  \displaystyle 
   \hskip50pt =b^2 q^{m-n}\mu^{+}R_{m-1,n}^{-}\bar{R}_{m,n}^{++}
    +x^2(1-b^2 q^{m-n})R_{m-1,n}^{+}\bar{R}_{m,n}, \\
 \smallskip
  \displaystyle 
   q^{\frac{1}{4}(-6m+2n-3)}x(1-q^{2m+1})R_{m-1,n}^{-}\bar{R}_{m+1,n-1} \\
 \smallskip
  \displaystyle 
   \hskip50pt =b^{-2}q^{-m+n}\nu R_{m,n-1}^{+}\bar{R}_{m,n}^{--}
    +x^2(1-b^{-2}q^{-m+n})R_{m,n-1}^{-}\bar{R}_{m,n}, \\
 \smallskip
  \displaystyle 
   q^{\frac{1}{4}(-2m+2n-1)}x(1-q^{2m+1})\bar{R}_{m-1,n-1}R_{m+1,n}^{-} \\
 \smallskip
  \displaystyle 
   \hskip50pt =b^{-2}q^{m+n+1}\nu R_{m,n-1}^{+}\bar{R}_{m,n}^{--}
    +x^2(1-b^{-2}q^{m+n+1})R_{m,n-1}^{-}\bar{R}_{m,n}, \\
 \smallskip
  \displaystyle 
  -q^{-\frac{1}{4}(2m+6n+3)}x(1-q^{2n+1})R_{m-1,n-1}\bar{R}_{m,n+1}^{-} \\
 \smallskip
  \displaystyle 
   \hskip50pt =b^2 q^{-m-n-1}\mu R_{m,n}^{--}\bar{R}_{m-1,n}^{+}
    +x^2(1-b^2 q^{-m-n-1})R_{m,n}\bar{R}_{m-1,n}^{-}, \\
 \smallskip
  \displaystyle 
  -q^{-\frac{1}{4}(2m+2n+1)}x(1-q^{2n+1})\bar{R}_{m,n-1}^{-}R_{m-1,n+1} \\
 \smallskip
  \displaystyle 
   \hskip50pt =b^2 q^{-m+n}\mu R_{m,n}^{--}\bar{R}_{m-1,n}^{+}
    +x^2(1-b^2 q^{-m+n})R_{m,n}\bar{R}_{m-1,n}^{-}, \\
 \smallskip
  \displaystyle 
   q^{-\frac{1}{4}(2m+2n+1)}x(1-q^{2m+1})R_{m+1,n-1}^{--}\bar{R}_{m-1,n}^{-} \\
 \smallskip
  \displaystyle 
   \hskip50pt =b^{-2}q^{m-n+2}\nu^{-}R_{m,n}\bar{R}_{m,n-1}^{---}
    +x^2(1-b^{-2}q^{m-n+2})R_{m,n}^{--}\bar{R}_{m,n-1}^{-}, \\
 \smallskip
  \displaystyle 
   q^{-\frac{1}{4}(6m+2n+3)}x(1-q^{2m+1})\bar{R}_{m+1,n}^{-}R_{m-1,n-1}^{--} \\
 \smallskip
  \displaystyle 
   \hskip50pt =b^{-2}q^{-m-n+1}\nu^{-}R_{m,n}\bar{R}_{m,n-1}^{---}
    +x^2(1-b^{-2}q^{-m-n+1})R_{m,n}^{--}\bar{R}_{m,n-1}^{-},
 \end{array}  \label{bi:bt:R}
\end{equation}
with 
\begin{equation}
\mu=(x+b^{-1})(x+q^{\frac{1}{2}}b^{-1}), \quad \nu=(x+b)(x+q^{-\frac{1}{2}}b), 
\end{equation}
where we denote as 
\begin{equation}
X^{\overbrace{\pm \cdots \pm}^j}=X^{\overbrace{\pm \cdots \pm}^j}(b)=
X(q^{\pm \frac{j}{2}}b). 
\end{equation}
\end{prop}
From the above discussion, now the proof of Theorem \ref{main} is reduced to that of Proposition \ref{bi:rel}. 

\medskip

It is possible to reduce the number of bilinear relations to be proved in (\ref{bi:bt:R}) by the following symmetry of $R_{m,n}^{(b)}(y|q)$. 

\begin{lem}\label{simple}
We have the relations for $m,n \in \BZ_{\ge 0}$ 
\begin{equation}
\begin{array}{c}
R_{n,m}^{(b^{-1})}(y|q^{-1})=R_{m,n}^{(b)}(y|q), \\
R_{n,m}^{(b^{-1})}=(-1)^{m(m+1)/2+n(n+1)/2}R_{m,n}^{(b)}. 
\end{array}
\end{equation}
\end{lem}

\noindent
{\it Proof.}~From (\ref{p-q:L}), it is easy to see that 
\begin{equation}
q_k^{(b)}(y|q)=p_k^{(b^{-1})}(y|q^{-1}), 
\end{equation}
which leads to the first relation of Lemma \ref{simple}. 
To verify the second relation, we introduce polynomials $\tilde{q}_k^{(b)}=\tilde{q}_k^{(b)}(y|q)$ by 
\begin{equation}
\sum_{k=0}^{\infty}\tilde{q}_k^{(b)}\lambda^k=
 \frac{(-q^{\frac{1}{4}}b^{-1}\lambda,
        -q^{\frac{3}{4}}b^{-1}\lambda;q)_{\infty}}
      {(q^{\frac{1}{4}}x\lambda,
        q^{\frac{3}{4}}x^{-1}\lambda;q)_{\infty}},
\quad \tilde{q}_k^{(b)}=0\ \mbox{for}\ k<0. \label{q+}
\end{equation}
Comparing the generating functions, we see that each $\tilde{q}_k$ is a linear combination of $q_j,~j=k,k-2,k-4,\cdots$. 
Therefore we can express $R_{m,n}^{(b)}$ for $m,n \in \BZ_{\ge 0}$ in terms of $p_k$ and $\tilde{q}_k$ as 
\begin{equation}
R_{m,n}^{(b)}=
 \left|
  \begin{array}{cccccccc}
   \tilde{q}_1^{(b)}        & \tilde{q}_0^{(b)}        & \cdots           & \tilde{q}_{-m+2}^{(b)}   &
   \tilde{q}_{-m+1}^{(b)}   & \cdots           & \tilde{q}_{-m-n+3}^{(b)} & \tilde{q}_{-m-n+2}^{(b)} \\
   \tilde{q}_3^{(b)}        & \tilde{q}_2^{(b)}        & \cdots           & \tilde{q}_{-m+4}^{(b)}   &
   \tilde{q}_{-m+3}^{(b)}   & \cdots           & \tilde{q}_{-m-n+5}^{(b)} & \tilde{q}_{-m-n+4}^{(b)} \\
   \vdots           & \vdots           & \ddots           & \vdots           &
   \vdots           & \ddots           & \vdots           & \vdots           \\
   \tilde{q}_{2m-1}^{(b)}   & \tilde{q}_{2m-2}^{(b)}   & \cdots           & \tilde{q}_m^{(b)}        &
   \tilde{q}_{m-1}^{(b)}    & \cdots           & \tilde{q}_{m-n+1}^{(b)}  & \tilde{q}_{m-n}^{(b)}    \\
   p_{n-m}^{(b)}    & p_{n-m+1}^{(b)}  & \cdots           & p_{n-1}^{(b)}    &
   p_n^{(b)}        & \cdots           & p_{2n-2}^{(b)}   & p_{2n-1}^{(b)}   \\
   \vdots           & \vdots           & \ddots           & \vdots           &
   \vdots           & \ddots           & \vdots           & \vdots           \\
   p_{-n-m+4}^{(b)} & p_{-n-m+5}^{(b)} & \cdots           & p_{-n+3}^{(b)}   &
   p_{-n+4}^{(b)}   & \cdots           & p_2^{(b)}        & p_3^{(b)}        \\
   p_{-n-m+2}^{(b)} & p_{-n-m+3}^{(b)} & \cdots           & p_{-n+1}^{(b)}   &
   p_{-n+2}^{(b)}   & \cdots           & p_0^{(b)}        & p_1^{(b)}
  \end{array}
 \right|. 
\end{equation}
Noticing that $\tilde{q}_k$ and $p_k$ are related as 
\begin{equation}
\tilde{q}_k^{(b)}(y|q)=(-1)^k p_k^{(b^{-1})}(y|q), 
\end{equation}
we obtain the second relation of Lemma \ref{simple}. \hfill\qed

\medskip

From the symmetries of $R_{m,n}^{(b)}(y|q)$ described by (\ref{neg:R}) and Lemma \ref{simple}, it is sufficient to prove the first two relations in (\ref{bi:bt:R}) for $m,n \in \BZ_{\ge 0}$, which are equivalent to 
\begin{equation}
-q^{\frac{1}{4}(2m-2n-1)}R_{m,n+1}^{+}\bar{R}_{m-1,n-1}
+q^{\frac{1}{4}(2m+2n+1)}\bar{R}_{m-1,n+1}R_{m,n-1}^{+}=
xR_{m-1,n}^{+}\bar{R}_{m,n}, 
\label{bi:1}
\end{equation}
\begin{equation}
-q^{\frac{1}{4}(2m-6n-3)}x(1-q^{2n+1})\bar{R}_{m-1,n+1}R_{m,n-1}^{+}=
b^2 q^{m-n}\mu^{+}R_{m-1,n}^{-}\bar{R}_{m,n}^{++}
+x^2(1-b^2 q^{m-n})R_{m-1,n}^{+}\bar{R}_{m,n},   \label{bi:2}
\end{equation}

In the following, we show that these bilinear relations are reduced to Jacobi's identity of determinants. 
Let $D$ be an $(m+n+1) \times (m+n+1)$ determinant and $\displaystyle D\left[\begin{array}{cccc} i_1 &
i_2 & \cdots & i_k \\ j_1 & j_2 & \cdots & j_k
       \end{array}
 \right]$ the minor which are obtained by deleting the rows with indices
$i_1,\cdots,i_k$ and the columns with indices $j_1,\cdots,j_k$.  Then we
have Jacobi's identity
\begin{equation}
D \cdot D\left[\begin{array}{cc}
           m &  m+1  \\
           1 & m+n+1
	  \end{array}
   \right]=
D\left[\begin{array}{c}
        m \\
        1
       \end{array}
  \right]
D\left[\begin{array}{c}
         m+1  \\
        m+n+1
       \end{array}
 \right]-
D\left[\begin{array}{c}
        m+1 \\
         1
       \end{array}
 \right]
D\left[\begin{array}{c}
          m   \\
        m+n+1
       \end{array}
 \right].         \label{Jacobi}
\end{equation}
We first choose proper determinants as $D$ ($D$ itself should be expressed in terms of $R_{m,n}^{(b)}$). 
Secondly, we construct such formulas that express the minor determinants by $R_{m,n}^{(b)}$. 
Then, Jacobi's identity yields bilinear equations for $R_{m,n}^{(b)}$ which are nothing but (\ref{bi:1}) and (\ref{bi:2}).

We have the following lemmas. 
\begin{lem}\label{dif:form:1}
We put 
\begin{equation}
D\equiv 
 \left|
  \begin{array}{ccccc}
   q^{-\frac{m+n-2}{2}}q^{\frac{1}{4}}x^{-1}q_1^+      & 
   \bar{q}_1      & \cdots & \bar{q}_{-m-n+3} & \bar{q}_{-m-n+2} \\
   q^{-\frac{m+n-4}{2}}q^{\frac{1}{4}}x^{-1}q_3^+      & 
   \bar{q}_3      & \cdots & \bar{q}_{-m-n+5} & \bar{q}_{-m-n+4} \\
   \vdots         & \vdots & \ddots & \vdots  & \vdots           \\
   q^{-\frac{n-m}{2}}  q^{\frac{1}{4}}x^{-1}q_{2m-1}^+ & 
   \bar{q}_{2m-1} & \cdots & \bar{q}_{m-n+1}  & \bar{q}_{m-n}    \\
   q^{-n}p_{n-m+1}^+       & \bar{p}_{n-m+2}  & \cdots & 
   \bar{p}_{2n}    & \bar{p}_{2n+1}   \\
   \vdots          & \vdots             & \ddots & 
   \vdots          & \vdots           \\
   q^{-1}p_{-n-m+3}^+ & \bar{p}_{-n-m+4} & \cdots & 
   \bar{p}_2        & \bar{p}_3        \\
   p_{-n-m+1}^+ & \bar{p}_{-n-m+2} & \cdots & 
   \bar{p}_0        & \bar{p}_1
  \end{array}
 \right|. 
\end{equation}
Then, we have 
\begin{equation}
\begin{array}{l}
 \displaystyle
  D=q^{\frac{1}{4}m^2-\frac{1}{4}n(n+5)}x^{-m}R_{m,n+1}^+, \quad 
 \displaystyle
   D\left[\begin{array}{c}
             m \\
             1
          \end{array}
    \right]
   =\bar{R}_{m-1,n+1}, \quad 
 \displaystyle
   D\left[\begin{array}{c}
            m+1 \\
             1
          \end{array}
    \right]
   =\bar{R}_{m,n}, \\
\medskip
 \displaystyle
   D\left[\begin{array}{c}
             m \\
           m+n+1
          \end{array}
    \right]
   =q^{\frac{1}{4}(m-1)^2-\frac{1}{4}(n-1)(n+4)-1}x^{-m+1}R_{m-1,n}^+, \\
\medskip
 \displaystyle
   D\left[\begin{array}{c}
            m+1  \\
           m+n+1
          \end{array}
    \right]
   =q^{\frac{1}{4}m^2-\frac{1}{4}(n-2)(n+3)-1}x^{-m}R_{m,n-1}^+, \quad 
 \displaystyle
   D\left[\begin{array}{cc}
             m &  m+1  \\
             1 & m+n+1
           \end{array}
     \right]
   =\bar{R}_{m-1,n-1}. 
\end{array}
\end{equation}
\end{lem}

\begin{lem}\label{dif:form:2}
Define $P_{j,k}^{[-m-n+j]}$ and $Q_{j,k}^{[-m-n+j]}$ by 
\begin{equation}
\begin{array}{l}
\displaystyle 
 P_{j,k}^{[-m-n+j]}=\prod_{i=1}^{m+n-j}(q^{-\frac{1}{4}-\frac{i}{2}}x)q^{\frac{1}{2}(m+n-j)\left(k-\frac{1}{2}(m+n-j)+\frac{1}{2}\right)}p_k^{[-m-n+j]}, \\
\displaystyle 
 Q_{j,k}^{[-m-n+j]}=q^{-\frac{1}{2}(m+n-j)k}q_k^{[-m-n+j]},
\end{array}
\end{equation}
where we denote 
\begin{equation}
X^{[j]}=X^{[j]}(x,b)=X(q^{\frac{j}{2}}x,q^{\frac{j}{2}}b). 
\end{equation}
Then, putting 
\begin{equation}
D\equiv
 \left|
  \begin{array}{ccccc}
   \widetilde{Q}_{0,1}^{[-m-n]++}        & Q_{1,1}^{[-m-n+1]}    & \cdots & 
   Q_{m+n-1,-m-n+3}^{[-1]}               & Q_{m+n,-m-n+2}^{[0]}  \\
   \widetilde{Q}_{0,3}^{[-m-n]++}        & Q_{1,3}^{[-m-n+1]}    & \cdots & 
   Q_{m+n-1,-m-n+5}^{[-1]}               & Q_{n+n,-m-n+4}^{[0]}  \\
   \vdots & \vdots & \ddots & \vdots & \vdots \\
   \widetilde{Q}_{0,2m-1}^{[-m-n]++}     & Q_{1,2m-1}^{[-m-n+1]} & \cdots & 
   Q_{m+n-1,m-n+1}^{[-1]}                & Q_{m+n,m-n}^{[0]}   \\
   \widehat{P}_{0,2n}^{[-m-n]++}         & P_{1,2n+1}^{[-m-n+1]} & \cdots & 
   P_{m+n-1,2n+1}^{[-1]}                 & P_{m+n,2n+1}^{[0]}      \\
   \vdots & \vdots & \ddots & \vdots     & \vdots \\
   \widehat{P}_{0,2}^{[-m-n]++}          & P_{1,3}^{[-m-n+1]}    & \cdots & 
   P_{m+n-1,3}^{[-1]}                    & P_{m+n,3}^{[0]}       \\
   \widehat{P}_{0,0}^{[-m-n]++}          & P_{1,1}^{[-m-n+1]}    & \cdots & 
   P_{m+n-1,1}^{[-1]}                    & P_{m+n,1}^{[0]}
 \end{array}
 \right|, 
\end{equation}
where 
\begin{equation}
\widehat{P}_{0,2k}^{[-m-n]++}=-\frac{P_{0,2k}^{[-m-n]++}}{1-q^{2k+1}}, \quad 
\widetilde{Q}_{0,2k-1}^{[-m-n]++}=\frac{Q_{0,2k-1}^{[-m-n]++}}{q^{m+n+1-2k}(1-q^{-m-n+2k}b^2)},
\end{equation}
we have 
\begin{equation}
 \begin{array}{l}
 \displaystyle
   D=(-1)^{n+1}
     \frac{\displaystyle \prod_{j=1}^{m+n}\mu^{[-m-n+j]}}
          {\displaystyle (q^{-\frac{1}{4}}b^{-2}x)^{m+n}
            \prod_{k=0}^n (1-q^{2k+1})
            \prod_{i=1}^m q^{m+n+1-2i}(1-q^{-m-n+2i}b^2)}R_{m,n}^{++}, \\
   \smallskip
  \displaystyle
   D\left[\begin{array}{c}
             m \\
             1
          \end{array}
    \right]
   =R_{m-1,n+1}, 
  \quad 
   D\left[\begin{array}{c}
            m+1 \\
             1
          \end{array}
    \right]
   =R_{m,n}, \\ \smallskip
  \displaystyle
   D\left[\begin{array}{c}
             m   \\
           m+n+1
          \end{array}
    \right]
   =(-1)^{n+1}x^{n+1}
     \frac{\displaystyle 
           q^{-\frac{1}{4}m^2+\frac{1}{4}n^2
              +\frac{1}{4}m-\frac{1}{2}n-\frac{3}{4}}
           \prod_{j=1}^{m+n-1}\mu^{[-m-n+j]}}
          {\displaystyle (q^{-\frac{1}{4}}b^{-2}x)^{m+n-1}
            \prod_{k=0}^n (1-q^{2k+1})
            \prod_{i=1}^{m-1} q^{m+n+1-2i}(1-q^{-m-n+2i}b^2)}
     {\underline R}_{m-1,n}^{+}, \\ 
   \smallskip
  \displaystyle
   D\left[\begin{array}{c}
            m+1  \\
           m+n+1
          \end{array}
    \right]
   =(-1)^n x^n
     \frac{\displaystyle 
           q^{-\frac{1}{4}m^2+\frac{1}{4}n^2-\frac{1}{4}m-n}
           \prod_{j=1}^{m+n-1}\mu^{[-m-n+j]}}
          {\displaystyle (q^{-\frac{1}{4}}b^{-2}x)^{m+n-1}
            \prod_{k=0}^{n-1} (1-q^{2k+1})
            \prod_{i=1}^m q^{m+n+1-2i}(1-q^{-m-n+2i}b^2)}
    {\underline R}_{m,n-1}^{+}, \\ 
   \smallskip
  \displaystyle
   D\left[\begin{array}{cc}
             m &  m+1  \\
             1 & m+n+1
           \end{array}
     \right]
   =q^{-\frac{1}{4}m^2+\frac{1}{4}n^2+\frac{1}{4}m-\frac{1}{2}n}x^n 
    {\underline R}_{m-1,n}^{-}. 
 \end{array}
\end{equation}
\end{lem}
It is easy to see that the bilinear relations (\ref{bi:1}) and (\ref{bi:2}) follow immediately from Jacobi's identity (\ref{Jacobi}) by using Lemma \ref{dif:form:1} and \ref{dif:form:2}, respectively. 
We give the proof of Lemma \ref{dif:form:1} and \ref{dif:form:2} in Appendix \ref{PL}. 
This completes the proof of our main result Theorem \ref{main}.

\section{Remarks \label{remarks}}
The $q$-P$_{\rm V}$ (\ref{q-P5}) admits the ultra-discrete limit~\cite{ultra}. 
The limiting procedure is the same as the case of $q$-P$_{\rm IV}$~\cite{KNY} and preserves the symmetry of the extended affine Weyl group of type $A_3^{(1)}$. 
Moreover, it is observed that $U_{m,n}=U_{m,n}(x,a)$ are polynomials in $x^{-1},a^{\pm 1}$ and $q^{\pm \frac{1}{2}}$ with positive coefficients. 
Then, the rational solutions of $q$-P$_{\rm V}$ (\ref{q-P5}) in Theorem \ref{main} are thought to survive after taking the ultra-discrete limit. 

It is known that the special polynomials associated with the rational solutions of Painlev\'e equations possess the mysterious combinatorial properties~\cite{Um1,Um2,Tane}. 
It is interesting problem to investigate whether the polynomials $U_{m,n}$ admit such properties. 

In~\cite{KNY}, it has been shown that the $q$-P$_{\rm IV}$ coincides with Sakai's Mul.6 system~\cite{elip}. 
As mentioned in Section \ref{Intro}, the $q$-P$_{\rm V}$ (\ref{q-P5}) has $\widetilde{W}(A_1^{(1)}\times A_3^{(1)})$ symmetry by the original construction. 
On the other hand, Sakai's Mul.5 system~\cite{elip}, which should be also regarded as a $q$-analogue of Painlev\'e V equation, admits the symmetry of $\widetilde{W}(A_4^{(1)})$. 
It might be an important problem to study the relationship between the equation (\ref{q-P5}) and Sakai's Mul.5 system. 

\medskip

\noindent 
{\bf Acknowledgment}\quad The author would like to thank Prof. M. Noumi, Prof. Y. Yamada and Prof. K. Kajiwara for fruitful discussions.

\appendix
\section{Table of $c_k,~\tilde{c}_k$ and $U_{m.n}$ \label{example}}
The polynomials $U_{m,n}(x,a)$. 
\begin{eqnarray*}
U_{0,0}&=&1, \\
U_{1,0}&=&1+q^{\frac{1}{2}}x^{-2}+a^{-1}q^{\frac{1}{2}}(1+q^{\frac{1}{2}})x^{-1},\\
U_{2,0}&=&1+q^{\frac{3}{2}}x^{-6}+a^{-1}(1+q^{\frac{1}{2}})(1+q+q^2)(x^{-1}+qx^{-5}) \\
       & &+q^{-\frac{1}{2}}(1+q+q^2)
            \left[1+a^{-2}q^{\frac{3}{2}}(1+q^{\frac{1}{2}})^2\right]
            (x^{-2}+q^{\frac{1}{2}}x^{-4}) \\
       & &+a^{-1}q^{\frac{1}{2}}(1+q^{\frac{3}{2}})
           \left[2(1+q^{\frac{1}{2}}+q)+a^{-2}q^2(1+q^{\frac{1}{2}})^2\right]x^{-3},
\end{eqnarray*}
\begin{eqnarray*}
U_{0,1}&=&1+aq^{\frac{1}{2}}(1+q^{\frac{1}{2}})x^{-1}+q^{\frac{1}{2}}x^{-2}, \\
U_{0,2}&=&1+q^{\frac{3}{2}}x^{-6}+a(1+q^{\frac{1}{2}})(1+q+q^2)(x^{-1}+qx^{-5}) \\
       & &+q^{-\frac{1}{2}}(1+q+q^2)
            \left[1+a^2q^{\frac{3}{2}}(1+q^{\frac{1}{2}})^2\right]
             (x^{-2}+q^{\frac{1}{2}}x^{-4}) \\
       & &+aq^{\frac{1}{2}}(1+q^{\frac{3}{2}})\left[2(1+q^{\frac{1}{2}}+q)+a^2q^2(1+q^{\frac{1}{2}})^2\right]x^{-3}, 
\end{eqnarray*}
\begin{eqnarray*}
U_{1,1}&=&1+qx^{-4}+(1+q^{\frac{1}{2}})(a+a^{-1})(x^{-1}+q^{\frac{1}{2}}x^{-3})
           +q^{-\frac{1}{2}}(1+q)(1+q^{\frac{1}{2}}+q)x^{-2}, \\
U_{1,2}&=&1+q^2x^{-8}
           +a^{-1}q^{-\frac{1}{2}}(1+q^{\frac{1}{2}})
            \left[1+a^2(1+q+q^2)\right](x^{-1}+q^{\frac{3}{2}}x^{-7}) \\
       & & +q^{-\frac{3}{2}}\left[a^2q^{\frac{3}{2}}(1+q^{\frac{1}{2}})^2(1+q+q^2)
          +(1+q^{\frac{1}{2}}+q+q^2)(1+q+q^{\frac{3}{2}}+q^2)\right](x^{-2}+qx^{-6})\\
       & &+a^{-1}q^{-1}(1+q^{\frac{3}{2}})
          \left[a^4q^2(1+q^{\frac{1}{2}})^2
               +a^2(1+q+q^2)(2+3q^{\frac{1}{2}}+2q)
               +(1+q^{\frac{1}{2}}+q)\right](x^{-3}+q^{\frac{1}{2}}x^{-5}) \\
       & &+q^{-1}(1+q^{\frac{1}{2}}+q+q^{\frac{3}{2}}+q^2)
          \left[a^2q^{\frac{1}{2}}(1+q^{\frac{1}{2}})(1+q)(1+q^{\frac{3}{2}})
                +2(1+q+q^2)\right]x^{-4}, \\
U_{2,1}&=&1+q^2x^{-8}
           +a^{-1}q^{-\frac{1}{2}}(1+q^{\frac{1}{2}})
            \left[a^2+(1+q+q^2)\right](x^{-1}+q^{\frac{3}{2}}x^{-7}) \\
       & &+a^{-2}q^{-\frac{3}{2}}
         \left[a^2(1+q^{\frac{1}{2}}+q+q^2)(1+q+q^{\frac{3}{2}}+q^2)
              +q^{\frac{3}{2}}(1+q^{\frac{1}{2}})^2(1+q+q^2)\right](x^{-2}+qx^{-6}) \\
       & &+a^{-3}q^{-1}(1+q^{\frac{3}{2}})
          \left[a^4(1+q^{\frac{1}{2}}+q)
               +a^2(1+q+q^2)(2+3q^{\frac{1}{2}}+2q)
               +q^2(1+q^{\frac{1}{2}})^2\right](x^{-3}+q^{\frac{1}{2}}x^{-5}) \\
       & &+a^{-2}q^{-1}(1+q^{\frac{1}{2}}+q+q^{\frac{3}{2}}+q^2)
          \left[2a^2(1+q+q^2)
            +q^{\frac{1}{2}}(1+q^{\frac{1}{2}})(1+q)(1+q^{\frac{3}{2}})\right]x^{-4}. 
\end{eqnarray*}

\medskip

\noindent
The factor $c_k$ and $\tilde{c_k}$. 
\[
\begin{array}{l}
c_0=1, \quad c_1=1, \\
c_2=1+q^{\frac{1}{2}}a^{-1}x^{-1}, \\
c_3=(1+q^{\frac{1}{2}}a^{-1}x^{-1})(1+qa^{-1}x^{-1})
    (1+q^{-\frac{1}{2}}ax^{-1}), \\
c_4=(1+q^{\frac{1}{2}}a^{-1}x^{-1})^2(1+qa^{-1}x^{-1})(1+q^{\frac{3}{2}}a^{-1}x^{-1})
    (1+q^{-\frac{1}{2}}ax^{-1})(1+q^{-1}ax^{-1}), \\
c_5=(1+q^{\frac{1}{2}}a^{-1}x^{-1})^2(1+qa^{-1}x^{-1})^2(1+q^{\frac{3}{2}}a^{-1}x^{-1})(1+q^2a^{-1}x^{-1}) \\
    \hskip25pt \times (1+q^{-\frac{1}{2}}ax^{-1})^2(1+q^{-1}ax^{-1})(1+q^{-\frac{3}{2}}ax^{-1}),
\end{array}
\]
\[
\begin{array}{l}
\tilde{c}_0=1, \quad \tilde{c}_1=1, \\
\tilde{c}_2=1+q^{-\frac{1}{2}}ax^{-1},  \\
\tilde{c}_3=(1+q^{-\frac{1}{2}}ax^{-1})(1+q^{-1}ax^{-1})
            (1+q^{\frac{1}{2}}a^{-1}x^{-1}), \\
\tilde{c}_4=(1+q^{-\frac{1}{2}}ax^{-1})^2(1+q^{-1}ax^{-1})(1+q^{-\frac{3}{2}}ax^{-1})
            (1+q^{\frac{1}{2}}a^{-1}x^{-1})(1+qa^{-1}x^{-1}), \\
\tilde{c}_5=(1+q^{-\frac{1}{2}}ax^{-1})^2(1+q^{-1}ax^{-1})^2(1+q^{-\frac{3}{2}}ax^{-1})(1+q^{-1}ax^{-2}) \\
           \hskip25pt \times (1+q^{\frac{1}{2}}a^{-1}x^{-1})^2(1+qa^{-1}x^{-1})(1+q^{\frac{3}{2}}a^{-1}x^{-1}). 
\end{array}
\]

\section{Proof of Lemma \ref{dif:form:1} and \ref{dif:form:2} \label{PL}}
We first note that the following contiguity relations hold, 
\begin{equation}
p_k^+ -q^{ \frac{k}{2}}\bar{p}_k=-q^{ \frac{1}{4}}x     p_{k-1}^+, \quad 
q_k^+ -q^{-\frac{k}{2}}\bar{q}_k=-q^{-\frac{1}{4}}x^{-1}q_{k-1}^+, 
\label{cont:1}
\end{equation}
\begin{equation}
p_k-q^{ \frac{k}{2}}{\underline p_k^-} =-q^{ \frac{3}{4}}x^{-1}p_{k-1}, \quad 
q_k-q^{-\frac{k}{2}}{\underline q_k^-} =-q^{-\frac{3}{4}}x     q_{k-1},
\label{cont:2}
\end{equation}
and 
\begin{equation}
 \begin{array}{l}
  (1-q^{k+1})p_{k+1}=
   q^{\frac{k}{2}+\frac{1}{4}}bx(b-b^{-1}){\underline p_k^+} 
   -q^{\frac{1}{4}}b^2x^{-1}\mu p_k^{++}, \\
  (1-q^{k+1}b^2)q_{k+1}=
   q^{\frac{1}{2}(k+1)}b(b^{-1}-b){\underline q_{k+1}^+}
   +q^{k+\frac{1}{4}}b^2x^{-1}\mu q_k^{++}, 
 \end{array}  \label{cont:3}
\end{equation}
which are easily derived from (\ref{p-q:(+-)}). 

Let us prove Lemma \ref{dif:form:1}. 
Noticing that $p_1=1$ and $p_k=0$ for $k<0$, we see that $R_{m,n}$ can be rewritten as 
\begin{equation}
R_{m,n}=
 \left|
  \begin{array}{cccccc}
   q_1        & q_0        & \cdots & q_{-m-n+3} & q_{-m-n+2} 
                    & q_{-m-n+1}  \\
   q_3        & q_2        & \cdots & q_{-m-n+5} & q_{-m-n+4} 
                    & q_{-m-n+3}  \\
   \vdots           & \vdots           & \ddots & \vdots           & \vdots 
                    & \vdots            \\
   q_{2m-1}   & q_{2m-2}   & \cdots & q_{m-n+1}  & q_{m-n}
                    & q_{m-n-1}   \\
   p_{n-m}    & p_{n-m+1}  & \cdots & p_{2n-2}   & p_{2n-1}
                    & p_{2n}      \\
   \vdots           & \vdots           & \ddots & \vdots           & \vdots
                    & \vdots            \\
   p_{-n-m+4} & p_{-n-m+5} & \cdots & p_2        & p_3
                    & p_4        \\
   p_{-n-m+2} & p_{-n-m+3} & \cdots & p_0        & p_1
                    & p_2        \\
   p_{-n-m}   & p_{-n-m+1} & \cdots & p_{-2}     & p_{-1}
                    & p_0        \\
  \end{array}
 \right|.    \label{edge:1}
\end{equation}
Adding the $(j-1)$-st column multiplied by $q^{\frac{1}{4}}x$ to the $j$-th column of $R_{m,n}^+$ for $j=m+n,m+n-1,\cdots 2$ and using (\ref{cont:1}), we get 
\begin{equation}
R_{m,n}^+ =q^{-\frac{1}{4}m^2+\frac{1}{4}(n-1)(n+4)}x^m
 \left|
  \begin{array}{ccccc}
   q^{-\frac{m+n-3}{2}}q^{\frac{1}{4}}x^{-1}q_1^+      & 
   \bar{q}_1      & \cdots & \bar{q}_{-m-n+4} & \bar{q}_{-m-n+3} \\
   q^{-\frac{m+n-5}{2}}q^{\frac{1}{4}}x^{-1}q_3^+      & 
   \bar{q}_3      & \cdots & \bar{q}_{-m-n+6} & \bar{q}_{-m-n+5} \\
   \vdots         & \vdots & \ddots & \vdots  & \vdots           \\
   q^{-\frac{n-m-1}{2}}q^{\frac{1}{4}}x^{-1}q_{2m-1}^+ & 
   \bar{q}_{2m-1} & \cdots & \bar{q}_{m-n+2}  & \bar{q}_{m-n+1}  \\
   q^{-n+1}p_{n-m}^+       & \bar{p}_{n-m+1}  & \cdots & 
   \bar{p}_{2n-2}  & \bar{p}_{2n-1}   \\
   \vdots          & \vdots             & \ddots & 
   \vdots          & \vdots           \\
   q^{-1}p_{-n-m+4}^+ & \bar{p}_{-n-m+5} & \cdots & 
   \bar{p}_2        & \bar{p}_3        \\
   p_{-n-m+2}^+    & \bar{p}_{-n-m+3} & \cdots & 
   \bar{p}_0        & \bar{p}_1
  \end{array}
 \right|.    \label{shift:1}
\end{equation}
From (\ref{edge:1}) and (\ref{shift:1}), we obtain Lemma \ref{dif:form:1}.

We next prove Lemma \ref{dif:form:2}. 
Adding the $(i+1)$-st column multiplied by $q^{\frac{1}{4}-(m+n-j)}x$ to the $i$-th column of $R_{m,n}$ for $i=1,2,\cdots,j,~j=m+n-1,m+n-2,\cdots,1$ and using (\ref{cont:2}), we get 
\begin{equation}
R_{m,n}=
 \left|
  \begin{array}{ccccc}
   Q_{1,1}^{[-m-n+1]}      & Q_{2,0}^{[-m-n+2]}    & \cdots & 
   Q_{m+n-1,-m-n+3}^{[-1]} & Q_{m+n,-m-n+2}^{[0]}  \\
   Q_{1,3}^{[-m-n+1]}      & Q_{2,2}^{[-m-n+2]}    & \cdots & 
   Q_{m+n-1,-m-n+5}^{[-1]} & Q_{n+n,-m-n+4}^{[0]}  \\
   \vdots & \vdots         & \ddots & \vdots & \vdots \\
   Q_{1,2m-1}^{[-m-n+1]}   & Q_{2,2m-2}^{[-m-n+2]} & \cdots & 
   Q_{m+n-1,m-n+1}^{[-1]}  & Q_{m+n,m-n}^{[0]}     \\
   P_{1,2n-1}^{[-m-n+1]}   & P_{2,2n-1}^{[-m-n+2]} & \cdots & 
   P_{m+n-1,2n-1}^{[-1]}   & P_{m+n,2n-1}^{[0]}    \\
   \vdots & \vdots         & \ddots & \vdots & \vdots \\
   P_{1,3}^{[-m-n+1]}      & P_{2,3}^{[-m-n+2]}    & \cdots & 
   P_{m+n-1,3}^{[-1]}      & P_{m+n,3}^{[0]}       \\
   P_{1,1}^{[-m-n+1]}      & P_{2,1}^{[-m-n+2]}    & \cdots & 
   P_{m+n-1,1}^{[-1]}      & P_{m+n,1}^{[0]}
 \end{array}
 \right|.   \label{shift:2}
\end{equation}
Noticing that $p_1=1$ and $p_k=0$ for $k<0$, we see that $R_{m,n}$ can be rewritten as 
\begin{equation}
R_{m,n}=
 \left|
  \begin{array}{ccccc}
   Q_{0,1}^{[-m-n]}        & Q_{1,0}^{[-m-n+1]}    & \cdots & 
   Q_{m+n-1,-m-n+2}^{[-1]} & Q_{m+n,-m-n+1}^{[0]}  \\
   Q_{0,3}^{[-m-n]}        & Q_{1,2}^{[-m-n+1]}    & \cdots & 
   Q_{m+n-1,-m-n+4}^{[-1]} & Q_{n+n,-m-n+3}^{[0]}  \\
   \vdots & \vdots         & \ddots & \vdots & \vdots \\
   Q_{0,2m-1}^{[-m-n]}     & Q_{1,2m-2}^{[-m-n+1]} & \cdots & 
   Q_{m+n-1,m-n}^{[-1]}    & Q_{m+n,m-n-1}^{[0]}   \\
   P_{0,2n}^{[-m-n]}       & P_{1,2n}^{[-m-n+1]}   & \cdots & 
   P_{m+n-1,2n}^{[-1]}     & P_{m+n,2n}^{[0]}      \\
   \vdots & \vdots         & \ddots & \vdots & \vdots \\
   P_{0,2}^{[-m-n]}        & P_{1,2}^{[-m-n+1]}    & \cdots & 
   P_{m+n-1,2}^{[-1]}      & P_{m+n,2}^{[0]}       \\
   P_{0,0}^{[-m-n]}        & P_{1,0}^{[-m-n+1]}    & \cdots & 
   P_{m+n-1,0}^{[-1]}      & P_{m+n,0}^{[0]}
 \end{array}
 \right|.   \label{edge:2}
\end{equation}
Then, adding the $j$-th column multiplied $\displaystyle \frac{q^{\frac{3}{4}+m+n-j}b^{-2}x(1-q^{-m-n+j}b^2)}{\mu^{[-m-n+j]}}$ to $(j+1)$-st column of $R_{m,n}^{++}$ for $j=m+n,m+n-1,\cdots,1$ and using (\ref{cont:3}), we obtain  
\begin{eqnarray}
R_{m,n}^{++}&=&(-1)^{n+1}
\frac{\displaystyle (q^{-\frac{1}{4}}b^{-2}x)^{m+n}\prod_{k=0}^n (1-q^{2k+1})
      \prod_{i=1}^m q^{m+n+1-2i}(1-q^{-m-n+2i}b^2)}
     {\displaystyle \prod_{j=1}^{m+n}\mu^{[-m-n+j]}} \\
&\times&
 \left|
  \begin{array}{ccccc}
   \widetilde{Q}_{0,1}^{[-m-n]++}    & Q_{1,1}^{[-m-n+1]}    & \cdots & 
   Q_{m+n-1,-m-n+3}^{[-1]}           & Q_{m+n,-m-n+2}^{[0]}  \\
   \widetilde{Q}_{0,3}^{[-m-n]++}    & Q_{1,3}^{[-m-n+1]}    & \cdots & 
   Q_{m+n-1,-m-n+5}^{[-1]}           & Q_{n+n,-m-n+4}^{[0]}  \\
   \vdots & \vdots & \ddots & \vdots & \vdots \\
   \widetilde{Q}_{0,2m-1}^{[-m-n]++} & Q_{1,2m-1}^{[-m-n+1]} & \cdots & 
   Q_{m+n-1,m-n+1}^{[-1]}            & Q_{m+n,m-n}^{[0]}   \\
   \widehat{P}_{0,2n}^{[-m-n]++}     & P_{1,2n+1}^{[-m-n+1]} & \cdots & 
   P_{m+n-1,2n+1}^{[-1]}             & P_{m+n,2n+1}^{[0]}      \\
   \vdots & \vdots & \ddots & \vdots & \vdots \\
   \widehat{P}_{0,2}^{[-m-n]++}      & P_{1,3}^{[-m-n+1]}    & \cdots & 
   P_{m+n-1,3}^{[-1]}                & P_{m+n,3}^{[0]}       \\
   \widehat{P}_{0,0}^{[-m-n]++}      & P_{1,1}^{[-m-n+1]}    & \cdots & 
   P_{m+n-1,1}^{[-1]}                & P_{m+n,1}^{[0]}
 \end{array}
 \right|,   \label{shift:3}
\end{eqnarray}
where we use the relations 
\begin{equation}
\begin{array}{l}
 \displaystyle
 {\underline P_{j,k}^{[-m-n+j+1]-}}=
 q^{-\frac{k}{2}+\frac{1}{4}+\frac{1}{2}(m+n-j)}x^{-1}P_{j,k}^{[-m-n+j]}, \\
 \displaystyle
 {\underline Q_{j,k}^{[-m-n+j+1]-}}=q^{\frac{k}{2}}Q_{j,k}^{[-m-n+j]}. 
\end{array}
\end{equation}
Lemma \ref{dif:form:2} follows from (\ref{shift:2}),(\ref{edge:2}) and (\ref{shift:3}).


\end{document}